# A nonextensive approach for understanding the van der Waals' equation

Lining Zheng and Jiulin Du[*]

Department of Physics, School of Science, Tianjin University, Tianjin 300072, China

**Abstract** In order to improve the teaching of the course of statistical physics in universities, in this article we introduce nonextensive statistics, a new statistical theory about complex systems. We study the two modification coefficients $a$ and $b$ in the van der Waals' equation in nonextensive statistics and thus understand the possible relation between the van der Waals' equation and the nonextensivity. We express these coefficients when a real gas is regarded as the nonextensive gas and then relate them to the $q$-parameter in nonextensive statistics. Furthermore, we derive the $q$-parameter of a real gas, which contain the intermolecular Lennard-Jones potential, but also strongly depend on the state parameter and molecular number of the gas. From the new statistical physics, we show that the nonextensivity plays a significant role in the coefficients of the van der Waals' equation if the gas contains limited number of molecules or the gas is regarded as a few-body system. This article is suitable to read for senior undergraduate students, graduate students and teachers who teach the course of statistical physics in universities.

E-mail: jldu@tju.edu.cn

## 1. Introduction

In the teaching of the course of statistical physics in universities, the equation of state of a gas is a very important content for undergraduate students to understand the statistical properties of a gas. Usually in the teaching of statistical physics, the equations of state of both an idea gas and a real gas are included, and the gases are naturally assumed to follow the traditional Boltzmann-Gibbs statistics. Under this traditional framework, as we know, the equation of state of a real gas can be approximately described by the famous van der Waals' equation. The standard form of van der Waals' equation is given [1] by

$$\left(p + \frac{a}{V^2}\right)(V - b) = NkT, \quad (1)$$

where $p$ is the gas pressure, $V$ is the volume, $k$ is the Boltzmann constant, $T$ is the temperature, $N$ is the molecular number, $a$ and $b$ are the modification coefficients of the pressure and the volume, respectively, introduced due to the intermolecular interactions in a real gas.

In the traditional statistical physics, there have been some ways to understand the equation of state of a real gas, such as the second virial coefficient and the Mayer theory, etc. The modification coefficients $a$ and $b$ are generally thought to originate from the interactions between particles in the gas and the temperature $T$ is a constant everywhere. If the intermolecular interactions can be expressed by Lennard-Jones potential,

$$\varphi(r) = \varphi_0 \left[\left(\frac{r_0}{r}\right)^{12} - 2\left(\frac{r_0}{r}\right)^{6}\right], \quad (2)$$

then by using the second virial coefficient, the above modification coefficients $a$ and $b$ in the van der Waals' equation can be expressed [1] as

$$a = \frac{2\pi}{3} N^2 \varphi_0 \, r_0^3, \quad b = \frac{2\pi}{3} N \, r_0^3, \quad (3)$$

where $\varphi_0$ is the interaction potential when the distance between two molecules is $r = r_0$

---

[*] Corresponding author

. However, in fact, the real gas can be in a non-thermal equilibrium state and in external fields. In this situation, we may need new statistical physics to understand the equations of state of complex systems including the van der Waals' equation. In fact, complexity has attracted one's attention in exploring to improve the teaching of university courses [2,3].

In recent years, a new statistical theory about the complex systems, called as nonextensive statistics, has been developed very well on the basis of the nonextensive entropy proposed by Tsallis in 1988 [4]. Nonextensive statistics can help us to understand the complex systems being in nonequilibrium states and having inter-particle interactions, and it has also obtained a lot of valuable applications in a variety of different science and technology fields, such as astrophysical and space plasmas [5-10], self-gravitating systems [11-14], chemical physics [15-18] and many other scientific researches [19-21]. Martinez et al discussed a generalization of the van der Waals' equation in nonextensive statistics [22]. The entropy in nonextensive statistics (also called Tsallis entropy or $q$-entropy) is given [4] as

$$S_q = \frac{k}{1-q}\left(\sum_i f_i^q - 1\right), \tag{4}$$

where $f_i$ is the probability that the system under consideration is in its $i$th configuration such that there is the normalization condition $\Sigma_i f_i = 1$, and $q$ is passitive and is the so-called nonextensive parameter whose deviation from 1 describes the degree of nonextensivity of the system. When we take $q \to 1$, the $q$-entropy can become Boltzmann entropy $S_B$,

$$\lim_{q \to 1} S_q = S_B = -k \sum_i f_i \ln f_i, \tag{5}$$

and thus the nonextensive statistics can recover to Boltzmann statistics.

Different from the tranditional Boltzmann entropy, the $q$-entropy (4) is nonextensive. The fundamental difference between the $q$-entropy $S_q$ and the B-G entropy $S_B$ lies in that, if the system is composed of two subsystems, $I$ and $II$, then the total $q$-entropy of the system is expressed by

$$S_q(I+II) = S_q(I) + S_q(II) + \frac{(1-q)}{k} S_q(I) S_q(II). \tag{6}$$

The entropy is nonextensive for the parameter $q \neq 1$. The extensivity is recovered only when we take the limit $q \to 1$.

The most important point in nonextensive statistics is the nonextensive parameter $q$, who's deviation from 1 is thought to be the degree of nonextensivity and also to stands for a deviation of nonextensive statistics from the traditional Boltzmann statistics. The $q$-parameter is closely related to the interaction properties between particles in a complex system. For example, when we consider a self-gravitating system, the $q$-parameter can be expressed by the gravitational potential $\varphi_g(r)$ and the temperature gradient [23] as

$$k\nabla T + (1-q)m\nabla \varphi_g(r) = 0, \tag{7}$$

where $m$ is mass of the particle. When we consider a complex plasma system, the $q$-parameter can be expressed by the electromagnetic field [24] and the temperature gradient as

$$k\nabla T = (1-q)e\nabla \varphi_c(r), \tag{8}$$

where $\varphi_c(r)$ is the Coulomb potential and $e$ is the electron charge. It is clear that $q=1$ if and only if $\nabla T = 0$. So the nonextensive statistics can describe the statistical property of the nonequilibrium complex system in the external fields.

In this paper, from nonextensive statistics we try to understand the van der Waals' equation of a real gas which is not in thermal equilibrium and which can be in an external field, to express the modification coefficients *a* and *b* in the equation using *q*-parameter in nonextensive statistics, and then to present the possible relation between the *q*-parameter and the intermolecular interactions in a nonequilibrium real gas. Of course, like many new theories in the history of science, nonextensive statistical mechanics is now still at a highly developed and controversial stage, where there are still many problems to be solved. So we hope to introduce this new statistical theory, nonextensive statistics, to undergraduate students and thus to improve the teaching of the course of statistical physics in universities.

The paper is organized as follows. In section 2, we first introduce the basic equations of state of systems in nonextensive statistics. In section 3, we study to express the modification coefficients *a* and *b* of the van der Waals equation using nonextensive statistics when a real gas is assumed to be the nonextensive gas, and then present the relation between the intermolecular interactions and the *q*-parameter. At the same time, we calculate the values of the *q*-parameter using the experimental data for the gases, $CO_2$, $N_2$, $O_2$, $H_2$ and He, which is as the understanding of the nonextensivity. Finally in section 5, we give the summary.

## 2. The equation of state in nonextensive statistics

One of important properties of nonextensive statistics is the nonextensivity of the *q*-entropy and the power-law distribution function. Namely, if a system is composed of two subsystems, I and II, then the total *q*-entropy of the system satisfies the equation (5). According to the maximum entropy principle, by using the *q*-entropy (1) we can get the probability distribution function in nonextensive statistics [25] to be the form,

$$f_i = \frac{1}{Z_q}\left[1-(1-q)\frac{\beta}{c_q}(\varepsilon_i - U_q)\right]^{\frac{1}{1-q}}. \quad (9)$$

This is a power-law distribution, where $\varepsilon_i$ is the energy in *i*th state, $\beta=1/kT$ is the Lagrange multiplier, $Z_q$ is the generalized canonical partition function given by

$$Z_q = \sum_i \left[1-(1-q)\frac{\beta}{c_q}(\varepsilon_i - U_q)\right]^{\frac{1}{1-q}}, \quad (10)$$

where $c_q = \sum_i f_i^q = Z_q^{1-q}$, and $U_q = c_q^{-1}\sum_i \varepsilon_i f_i^q$ is the generalized internal energy. If we take the limit $q \to 1$, the power-law *q*-distribution function (9) recovers the famous Boltzmann distribution function,

$$f_i = \frac{1}{Z}\exp\left[-\beta(\varepsilon_i - U)\right], \quad (11)$$

and $Z_q$ recovers the standard partition function in Boltzmann-Gibbs statistics,

$$Z = \sum_i \exp\left[-\beta(\varepsilon_i - U)\right]. \quad (12)$$

According the above equations, one can find the relation between the *q*-entropy and the generalized canonical partition function,

$$S_q = \frac{k}{1-q}(c_q - 1) = \frac{k}{1-q}(Z_q^{1-q} - 1) \quad (13)$$

with $c_q = Z_q^{1-q}$. And in nonextensive statistics, the pressure can be defined [26] as

$$P_q = \frac{c_q T}{1+(1-q)S_q/k} \frac{\partial S_q}{\partial V} \tag{14}$$

with the volume $V$ of the system. It has been proved that the pressure in nonextensive statistics can be expressed as

$$P_q = kTZ_q^{1-q} \frac{\partial}{\partial V} \ln Z_q = \frac{kTc_q}{1-q} \frac{\partial}{\partial V} \ln c_q. \tag{15}$$

As an example, for the classical $D$-dimension nonextensive gas with particle number $N$, the generalized canonical partition function is known [27, 28] as

$$Z_q = \frac{\Gamma[(2-q)/(1-q)] c_q^{DN/2}}{\Gamma[(2-q)/(1-q)+DN/2]} \frac{V^N}{N!h^{DN}} \left(\frac{2\pi mkT}{1-q}\right)^{DN/2} \left[1+(1-q)\frac{DN}{2}\right]^{1/(1-q)+DN/2} \tag{16}$$

Using (15) and (16), we can therefore derive the equation of state of the nonextensive gas, namely,

$$P_q V = NkTZ_q^{1-q}. \tag{17}$$

## 3. The nonextensive parameter and van der Waals' equation

### 3.1 *The modification coefficients a and b from nonextensive statistics*

We now study the modification coefficients $a$ and $b$ in the van der Waals' equation if we can regard a nonextensive gas as a real gas. We first give a 3-dimension nonextensive gas model with $N$ particles: the Hamiltonian reads $H=\sum_{i=1}^{N}\mathbf{p}_i^2/2m$, where $\mathbf{p}_i$ is the momentum of $i$th particle ($i$=1,2,..., $N$). For this gas model, the probability distribution function in nonextensive statistics is expressed [26-28] as

$$f(\{\mathbf{p}_i\}) = \frac{1}{Z_q}\left[1-(1-q)\frac{\beta}{c_q}\left(\sum_i^N \frac{\mathbf{p}_i^2}{2m} - U_q\right)\right]^{\frac{1}{1-q}}, \tag{18}$$

which can define the steady undisturbed state of a complex system with inter-particle interactions if the nonextensivity introduced by the particle interactions are reflected by the nonextensive parameter $q\neq 1$, where one can write the generalized canonical partition function [28] that

$$Z_q^{1-q} = \left[\frac{\Gamma[1/(1-q)]}{\Gamma[1/(1-q)+3N/2]} \frac{V^N}{N!h^{3N}}\left(\frac{2\pi mkT}{1-q}\right)^{3N/2}\right]^{[2(1-q)]/[2-(1-q)3N]}$$
$$\times \left[1+(1-q)\frac{3N}{2}\right]^{[2q+(1-q)3N]/[2-(1-q)3N]} \tag{19}$$

and $0 < q < 1$. The nonextensive gas model has been applied to study many physical quantities in real gases [6,14], such as the Joule coefficient, the second virial coefficient and so on. Using the state equation (17) of nonextensive gas we have immediately that

$$\left(\frac{\partial P_q}{\partial T}\right)_V = \frac{Nk}{V}\frac{2c_q}{2-(1-q)3N}. \tag{20}$$

And thus we can obtain,

$$T\left(\frac{\partial P_q}{\partial T}\right)_V - P_q = \frac{(1-q)3N}{2-(1-q)3N}P_q. \tag{21}$$

On the other hand, from the van der Waals' equation (1) of a real gas we have that

$$\left(\frac{\partial p}{\partial T}\right)_V = \frac{Nk}{V-b}, \text{ and } T\left(\frac{\partial p}{\partial T}\right)_V - p = \frac{a}{V^2}. \tag{22}$$

If we can regard the nonextensive gas as a real gas so as to understand the modification coefficients $a$ and $b$ in the van der Waals' equation, i.e., if we let $P_q = p$, then from Eqs.(20), (21) and (22) we obtain that

$$a = \frac{(1-q)3N}{2-(1-q)3N}pV^2, \tag{23}$$

$$b = V - \frac{NkT}{p}\left[1-(1-q)\frac{3N}{2}\right]. \tag{24}$$

In the two equations (23) and (24), we show that the modification coefficients $a$ and $b$ in the van der Waals' equation of a real gas can be related to the nonextensive parameter $q$ in nonextensive statistics in such a way if a real gas can be regarded as a nonextensive gas, and simultaneously they also depend on the state parameters, such as volume, pressure and temperature. In other words, the coefficients $a$ and $b$ are not constant, but vary according to physical properties of a gas, in which the nonextensivity plays a role. If we take $q=1$, we have $a=0$ and $(V-b)p = NkT$, representing the situation that the interaction has a modification only to the volume but has no modification to the pressure.

3.2 *The nonextensive parameter understanding a real gas*

The equations (23) and (24) show us an important revelation that the link between the modification coefficients and the degree of nonextensivity of the gas depends strongly on the total molecule number $N$. If $N$ is very large, or if we take $N \to \infty$, we can see that $a \approx -pV^2$ and $b \approx V-NkT/p$, and so the nonextensivity has no effect on the van der Waals' equation. Only when the gas contains limited number of molecules or the gas is a few-body system, the nonextensivity plays a significant role in the modification coefficients $a$ and $b$ of the van der Waals' equation.

Furthermore, using Eq.(3), Eqs.(23) and (24) we can obtain two new expressions of the nonextensive parameter $q$ for a real gas,

$$1-q = 2\left(\frac{9pV^2}{2\pi N\varphi_0 r_0^3} + 3N\right)^{-1}, \tag{25}$$

which contain the Lennard-Jones potential, and

$$1-q = \frac{2}{3N}\left[1 - \frac{p}{NkT}\left(V - \frac{2\pi}{3}Nr_0^3\right)\right], \tag{26}$$

which contain the temperature of the gas. Thus we find that the nonextensive parameter $q$ of a real gas can be determined by the several different factors, including the pressure, the temperature, the volume, the molecular number and the interaction property between molecules. If we take $N \to \infty$, but $N/V$ is limited, we have $q \to 1$ and thus the system becomes extensive, which reaches the conclusion that only a few-body system or so-called small system may be nonextensive, but a real gas should generally be extensive because $N$ is very large.

3.3 *Experimental data of the nonextensive parameter of real gases*

As examples, we calculated 1-$q$ value by using the experimental data for five real gases in order to test the new theory and the nonextensivity of a real gas. In table 1, we listed the experimental values of the modification coefficients $a$ and $b$ in the van der Waals' equation for the five different gases [1], $CO_2$, $N_2$, $O_2$, $H_2$ and He. These experimental data are obtained all for one mol gas and at the temperature $T$=273.15K and the volume $V$=5.5 ×$10^{-4}m^3$. The corresponding pressures of the gases are also listed in the table.

Using these experimental data and the equations (23) and (24), we can calculate the parameter (1-$q$), which therefore is as a test of the new theory and the nonextensivity of a real gase. The values of (1-$q$) for the five gases are given in table 1.

Because the deviation of $q$ from 1 descibes the degree of nonextensivity of a system and only when $q$=1 the system becomes extensive, from table 1 we can see that the nonextensive parameter (1-$q$) is so small that these real gases are generally extensive. Therefore, As we imagine, they can be well described by the traditional statistical mechanics.

Table 1. Calculations from the experimental values of five real gases

| Gas | $a$ /(Pa $m^6$) | $b$ /($10^{-5}m^3$) | $p$/($10^6$Pa) | (1-$q$)/$10^{-25}$ |
|---|---|---|---|---|
| $CO_2$ | 0.3640 | 4.267 | 3.274 | 2.975 |
| $N_2$ | 0.1408 | 3.913 | 3.981 | 1.159 |
| $O_2$ | 0.1378 | 3.183 | 3.928 | 1.150 |
| $H_2$ | 0.02476 | 2.661 | 4.258 | 0.209 |
| He | 0.003456 | 2.370 | 4.305 | 0.029 |

## 4. Summary

In summary, we have studied the modification coefficients $a$ and $b$ in the van der Waals' equation of a real gas if the real gas can be regarded as a nonextensive gas in nonextensive statistics. We have derived expressions of $a$ and $b$ for the nonextensive gas, Eqs.(23) and (24), in which we show that $a$ and $b$ relate the nonextensive parameter $q$, but simultaneously they also depend on the state parameters of the gas, such as molecular number, volume, pressure and temperature. In other wards, the coefficients $a$ and $b$ are not constant, but vary according to physical properties of a gas, in which the nonextensivity plays a role. Therefore, to some extent, the parameter $q$ can be applied to describe the intermolecular interactions in a real gas.

Firstly, the equations (23) and (24) show us an important revelation that the link between the modification coefficients and the degree of nonextensivity of the gas depends strongly on the total molecule number $N$. If $N$ is very large, we can see that $a \approx -pV^2$ and $b \approx V-NkT/p$, and so the nonextensivity has no effect on the van der Waals' equation. Only when the gas contains limited number of molecules or the gas is a few-body system, the nonextensivity plays a significant role in the modification coefficients $a$ and $b$ of the van der Waals' equation.

Secondly, we obtained two new expressions of the nonextensive parameter $q$ for a real gas, Eqs.(24) and (25), which are associated with the Lennard-Jones potential between molecules. We can determine the parameter $q$ by using these two expressions. If we take the molecular number $N\rightarrow\infty$, we have $q\rightarrow 1$ and thus the system becomes extensive, which reaches the conclusion that only a few-body system or so-called small system may be nonextensive, but a real gas should generally be extensive because $N$ is very large.

Furthermore, we used the experimental data of the modification coefficients *a* and *b* in the van der Waals' equations for five real gases, $CO_2$, $N_2$, $O_2$, $H_2$ and He, and their state parameters to determine values of the nonextensive parameter *q* of these gases, so it can be as a experimental test of the above conclusions for the van der Waals' equation of the nonextensive gas. The results prove that the degree of the nonextensivity is very low for a real gas.

Finally, we hope that this article introduce this new statistical theory, nonextensive statistics, to undergraduate students and teachers who teach the couse of statistical physics, and then can improve the teaching of statistical physics in universities.

**Acknowledgements**

This work was supported by the National Natural Science Foundation of China under grant No.11175128.

**References**

[1] Wang Zhicheng 2008 *Thermodynamics and Statistical Physics* (Beijing: Higher Education Press)
[2] Holovatch Y, Kenna R and Thurner S 2017 *Eur. J. Phys.* **38** 023002
[3] Perc M 2018 *Eur. J. Phys.* **39** 014001
[4] Tsallis C 2009 *Introduction to Nonextensive Statistical Mechanics: Approaching a Complex World*, (New York: Springer)
[5] Liu Z, Liu L and Du J 2009 *Phys. Plasmas* **16** 072111
[6] Gong J, Liu Z and Du J 2012 *Phys. Plasmas* **19** 083706
  Du J 2013 *Phys. Plasmas* **20** 092901
[7] Yu H and Du J 2016 *EPL* **116** 60005
[8] Abbasi Z E and Esfandyari-Kalejahi A 2016 *Phys. Plasmas* **23** 073112
  Du J, Guo R, Liu Z and Du S 2018 *Contrib. Plasma Phys.* doi: 10.1002/ctpp.201800046
[9] Ghosh D K, Mandal G, Chatterjee P and Ghosh U N 2013 *IEEE Trans. Plasma Sci.* **41** 1600
[10] Sahu B and Tribeche M 2012 *Astrophys. Space Sci.* **338** 259
[11] Du J 2007 *Astrophys. Space Sci.* **312** 47
  Du J 2004 *EPL* **67** 893
[12] Cardone V F, Leubner M P and Del Popolo A 2011 *Mon. Not. R. Astron. Soc.* **414** 2265
[13] Zheng Y 2013 *EPL* **101** 29002
[14] Zheng Y 2013 *EPL* **102** 10009
[15] Du J 2012 *Physica A* **391** 1718
[16] Yin C and Du J 2014 *Physica A* **395** 416
[17] Carvalho-Silva V H, Coutinho N D and Aquilanti V 2016 *AIP Conf. Proc.* **1790** 020006
[18] Zhou Y and Yin C 2015 *Physica A* **417** 267
  Zhou Y and Yin C 2016 *Int. J. Mod. Phys. B* **30** 1650095
[19] Zheng Y and Du J 2007 *Int. J. Mod. Phys. B* **21** 947
  Zheng Y 2013 Physica A **392** 2487
[20] Kaniadakis G 2013 *Entropy* **15** 3983
[21] Zhou Y and Du J 2016 *Fluct. Noise Lett.* **15** 1650001
[22] Martinez S, Pennini F and Plastino A 2001 *Phys. Lett. A* **282**,263
[23] Du J 2004 *Europhys. Lett.* **67** 893
[24] Du J 2004 *Phys.Lett.A* **329** 262
[25] Tsallis C, Mendes R S and Plastino A R 1998 Physica A **261** 534
[26] Abe S, Martinez S, Pennini F and Plastino A 2001 *Phys. Lett. A* **281** 126
[27] Abe S 1999 *Phys. Lett. A* **263** 424
[28] Du J 2004 *Physica A* **335** 107